\documentclass{article}
\usepackage{spconf,amsmath,epsfig}

\usepackage{graphicx}
\usepackage{amsmath}
\usepackage{diagbox,multirow}
\usepackage{makecell}

\usepackage{float}

\let\OLDthebibliography\thebibliography
\renewcommand\thebibliography[1]{
  \OLDthebibliography{#1}
  \setlength{\parskip}{0pt}
  \setlength{\itemsep}{0pt plus 0.3ex}
}

\begin{document}\sloppy

% Example definitions.
% --------------------
\def\x{{\mathbf x}}
\def\L{{\cal L}}

% Title.
% ------
\title{ConvNeXt Based Neural Network for Audio Anti-Spoofing}
%
% Single address.
% ---------------
\name{Qiaowei Ma, Jinghui Zhong, Yitao Yang, Weiheng Liu, Ying Gao and Wing W.Y. Ng}
%Address and e-mail should NOT be added in the submission paper. They should be present only in the camera ready paper. 
\address{South China University of Technology\\
jinghuizhong@scut.edu.cn}

\maketitle

\begin{abstract}
With the rapid development of speech conversion and speech synthesis algorithms, automatic speaker verification (ASV) systems are vulnerable to spoofing attacks. In recent years, researchers had proposed a number of anti-spoofing methods based on hand-crafted features. However, using hand-crafted features rather than raw waveform will lose implicit information for anti-spoofing. Inspired by the promising performance of ConvNeXt in image classification tasks, we revise the ConvNeXt network architecture and propose a lightweight end-to-end anti-spoofing model. By integrating with the channel attention block and using the focal loss function, the proposed model can focus on the most informative sub-bands of speech representations and the difficult samples that are hard to classify. Experiments show that our proposed system could achieve an equal error rate of 0.64\% and min-tDCF of 0.0187 for the ASVSpoof 2019 LA evaluation dataset, which outperforms the state-of-the-art systems.
\end{abstract}
\section{Introduction}
\label{sec:intro}

Automatic speaker verification (ASV) systems use the speaker's voiceprint information to identify specific person, and have been widely used in many real life applications. However, with the rapid development of the text to speech (TTS) \cite{prenger2019waveglow} and voice conversion (VC) \cite{tanaka2019atts2s} algorithms, the ASV systems are vulnerable to the spoof attacks.

In order to promote the research on spoof attack detection, the bi-annual anti-spoofing challenge was first held in 2015 \cite{wu2015asvspoof}. Since then, many spoof attack detection methods had been proposed. And these works mostly focus on feature engineering and network architecture designing. For the feature engineering, hand-crafted features such as phase information \cite{sanchez2015toward}, octave spectra information \cite{yang2019extraction}, Constant Q Cepstral Coefficients (CQCC) \cite{todisco2017constant}, CQT-based modified group delay feature \cite{cheng2019replay} and genuinized feature that transformed from the given log power spectrum (LPS) \cite{wu20c_interspeech} were proposed to enhance the effectiveness of the anti-spoofing system. For the network architecture design, the residual network and its variant \cite{cheng2019replay,alzantot19_interspeech} were used to detect the spoof attacks, Wu et al.\cite{wu20c_interspeech} use a light convolutional neural network(LCNN) against spoofing attacks. Recently Li et al.\cite{li2021replay} use a Res2Net architecture for replay and synthetic speech detection. Res2Net is a multiple-feature scale network architecture. It split the features maps within one block into multiple channel groups and designs a residual-like connection across different channel groups, which could increase the possible receptible fields and enhance the anti-spoofing capacity of the model. Although the above residual network architecture or its variant achieves competitive performance for spoof attack detection, these single systems still lack generalizability to unseen spoofing attacks. Ensemble systems that use different hand-crafted features would improve the anti-spoofing performance, but it will increase the model's complexity.

Extracting hand-crafted features would lose some useful information for detecting spoof attacks, for example, the log power spectrum of the CQT feature, does not have the phase information of the signal. Therefore, some researchers proposed the anti-spoofing model that directly uses the raw waveform as the network's input to capture the spoof cues of the spoof audio \cite{hua2021towards,jung2022aasist,hansen22_interspeech}. The results of these end-to-end systems are better than many anti-spoofing systems that use hand-crafted features.

Recently, Liu et al. modernized a standard ConvNet (ResNet \cite{he2016deep}) towards the design of a hierarchical vision transformer and proposed a pure ConvNet model called ConvNeXt. \cite{liu2022convnet}. ConvNeXt had achieved great success in some vision tasks. Its classification performance demonstrated ConvNeXt's powerful feature extraction capability. Inspired by the success of the residual network and its variant used for spoof attack detection and the promising performance of the end-to-end spoof attack detection, we design an end-to-end anti-spoofing neural network by revising ConvNeXt's network architecture. The major contributions of this paper are as follows:
\begin{itemize}
    \item We proposed a new fully end-to-end anti-spoofing neural network (named CNBNN) by revising the ConvNeXt's network architecture, which achieved outstanding performance and outperform state-of-the-art systems.
    \item We revise the efficient channel attention (ECA) module from \cite{9156697} and integrate it into the proposed model. Such a channel attention module could make the model focus on the most informative sub-bands of speech representations to improve the detection performance of the system.
    \item The proposed best single system could achieved an equal error rate (EER) of 0.64\% and min-tDCF of 0.0187 on the evaluation dataset of ASVSpoof 2019 LA dataset, which demonstrate the model's capacity for anti-spoofing.
\end{itemize}

The rest of this paper is organized as follows: Section 2 illustrates the proposed methods. Experiment setup, results and analysis are discussed in Section 3 and 4, respectively. We conclude this work in Section 5.

\section{Proposed methods}
This section will first present a general framework of our proposed system and then introduce the model's res2net style block and Modified ECA layer in detail. Finally, we will introduce the focal loss used in our model.
\subsection{ConvNeXt based neural network architecture}
\begin{figure*}[ht]
    \centerline{\includegraphics[scale=0.2705]{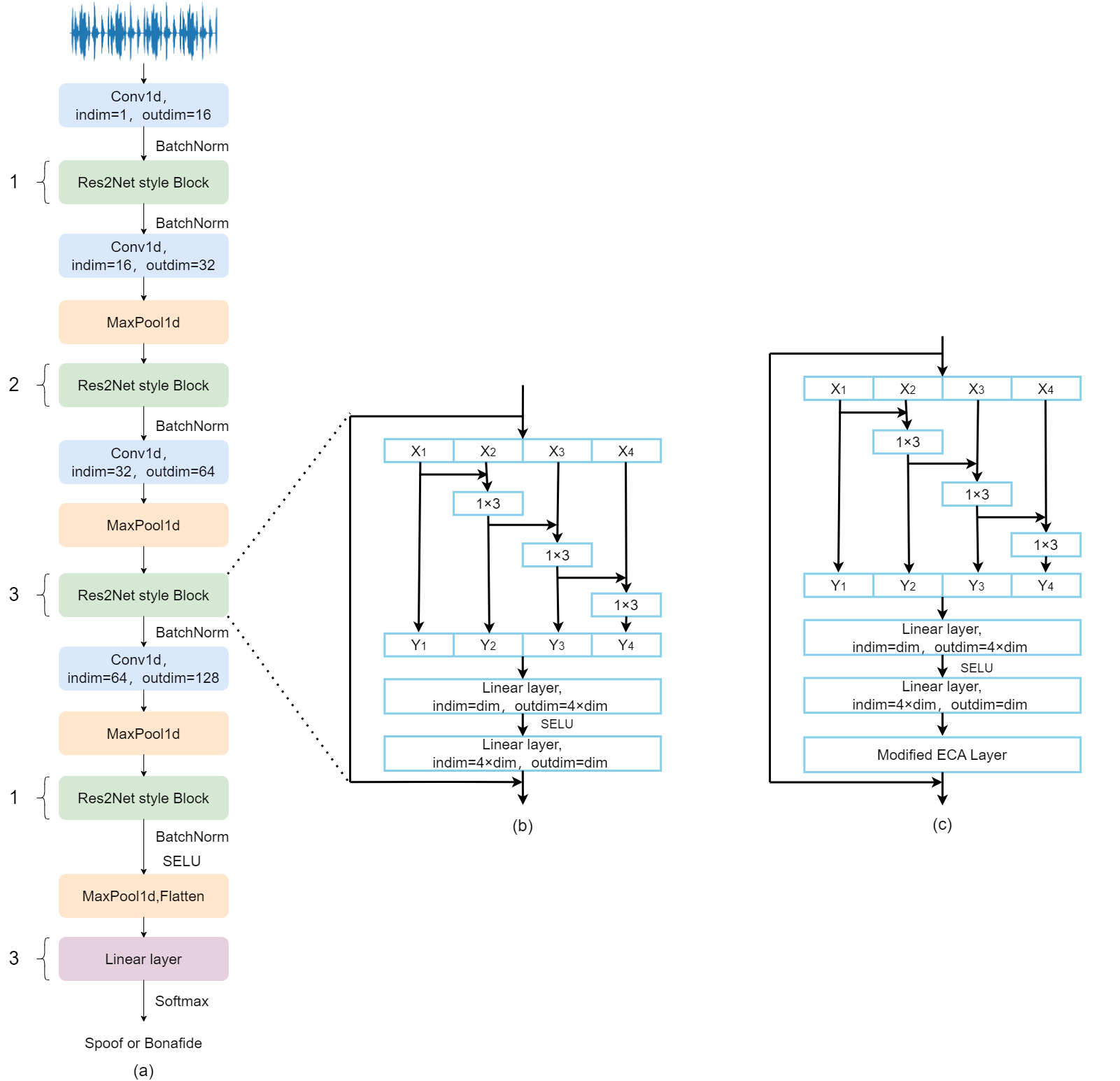}}
    \caption{ConvNeXt based neural network. (a)an overall architecture of the proposed model. (b)the detail of the Res2net style Block. (c)Res2net style Block integrated with a channel attention module(Modified ECA layer)}
    \label{fig:convnext_architecture}
\end{figure*}
As illustrated in figure \ref{fig:convnext_architecture}.(a), we adapt the ConvNeXt network architecture from \cite{liu2022convnet}. We revise the network to make it suitable for the input raw waveform, thus we can train the neural network end-to-end. Compared with the original network architecture, we remove the Stochastic Depth \cite{huang2017densely}, which we find does not improve the performance of the anti-spoofing system. For the normalization method, we replace layer normalization with batch normalization and achieve better performance for anti-spoofing. We assume that the network should be relatively shallower because a deeper network extract deeper features which towards higher level semantic information, but it may not suitable to represent the subtle forgery artifacts from the spoof attacks. So we change the different compute stage ratios 3:3:9:3 to 1:2:3:1 to make the network shallower. We use a max pooling layer with a large kernel size of 9 to downsample the feature map instead of just using the convolution layer except the stem layer. And we set the channel of each compute stage to be (16,32,64,128). For the activation function, we replace GELU with SELU, which achieved better performance for anti-spoofing. We revise the ResNet style Block in the network according to the Res2Net block and then design a res2net style block, thus such design could make the model have a multi-scale receptive field and then achieve better performance for detecting the spoof artifact. And we integrate a channel attention module that modified from the ECA layer with it to enhance the model's robustness and generalization capacity.
\subsection{Res2Net style block}
As illustrated in figure \ref{fig:convnext_architecture}.(b), the Res2Net style block use a reverted bottleneck. It means that the hidden dimension of the MLP block is four times wider than the input dimension. To be specify, the input feature maps were first through a res2net style convolutional layer, it splits the input feature maps by the channel dimension into 4 subsets, denoted by ${{X}_{i}}$, where $i\in \text{ }\!\!\{\!\!\text{ 1,2,3,4 }\!\!\}\!\!\text{ }$. Except for ${{X}_{1}}$, each ${{X}_{i}}$ is processed by a 1 × 3 convolutional filter ${{K}_{i}}$. Starting from i = 2, ${{X}_{i}}$ is added with the output of ${{K}_{i-1}}$ before being fed into ${{K}_{i}}$(). This process can be formulated as Eq. \ref{eq1}:
\begin{equation}
    \label{eq1}
    Y_i=
    \begin{cases}
        X_i,&i= 1 \\
        K_i(X_i+Y_{i-1}),&2\le i \le 4
    \end{cases}
\end{equation}
In the last of the res2net style block, two linear layers is used to enlarge and fuse information between each channel. The block that integrated with the channel attention module (MECA) is illustrated in figure \ref{fig:convnext_architecture}.(c).
\subsection{Modified ECA layer}
To further enhance the robustness and generalizability of the model, we use a channel attention module that revises from the original ECA layer to make it suitable for the input raw waveform. ECA layer is a channel attention module that revises from the SE Block \cite{hu2018squeeze}, it is assumed that dimensionality reduction that used in SE block is unnecessary to capture the dependencies between each channel. So in ECA layer, it avoids dimensionality reduction and captures cross-channel interaction in an efficient way.
\begin{figure}[H]
\centerline{\includegraphics[scale=0.4]{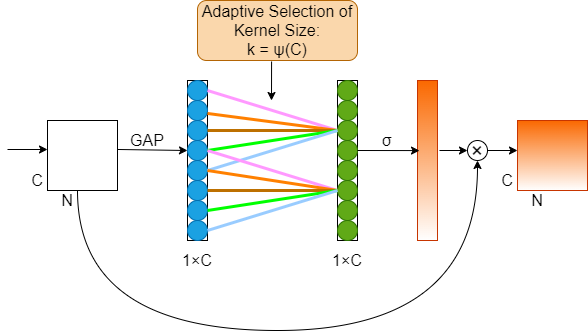}}
\caption{Diagram of the modified efficient channel attention (MECA) module. Given the aggregated features obtained by global average pooling (GAP), Modified ECA generates channel weights by peforming a fast 1D convolution of size k, where k is adaptively determined via a mapping of channel dimension C.}
\label{fig:meca}
\end{figure}
As illustrated in figure \ref{fig:meca}, given a feature map $x\in {{R}^{C\times N}}$, where $C$ represents the numbers of the channel and $N$ represents the numbers of the sampling points of the input raw waveform. A channel-wise global average pooling (GAP) without dimensionality reduction is used to gain the representation of each channel $y$, and then a fast 1D convolution of size $k$ is used to capture local cross-channel interaction. this process can be represented by Eq. \ref{eq2}:
\begin{equation}
    \label{eq2}
    y = \sigma(Conv1d_k(\frac{x}{N}))
\end{equation}
where $k$ represents how many neighbors participate in attention prediction of one channel. And in this work, we use the method that used in \cite{9156697} to adaptively select the kernel size, which is proportional to channel dimension. The way to compute $k$ is represented by Eq. \ref{eq3}:
\begin{equation}
    \label{eq3}
    k = \psi(C) = \bigg|\frac{log_2(C)}{\gamma}+\frac{b}{\gamma}\bigg|_{odd},
\end{equation}
where $|t|_{odd}$ indicates the nearest odd number of t. $\gamma$ and $b$ are the function parameters and are set to 2 and 1 respectively. When we compute the weight of each channel $y$, we use $y$ to reweight the original feature map. Such a channel attention module could make the model learn to emphasize important features and suppress useless features. Thus the model could focus on the most informative sub-bands of speech representations and have a better performance for spoof attack detection.
\subsection{Focal loss}
Since there exist different kinds of spoof attacks in the training set and different kinds of attacks have different degrees of difficulty for the model to classify. Also, there exists an imbalanced class distribution on the ASVSpoof 2019 dataset, the number of the bonafide speech is much less than the spoof speech. So in this case, We use focal loss \cite{lin2017focal} to make the model focus more on the hard-to-classify samples in spoof attacks and also deal with the data imbalanced during the training phase. During training, the samples that are hard to classify would be given more loss, and the samples that are easy to classify would be given less loss. The focal loss could be formulated by Eq. \ref{eq4}:
\begin{equation}
    \label{eq4}
    y = -\alpha_t(1-p_t)^ylog(p_t).
\end{equation}
In the above, $\alpha_t$ is loss weight that assign to the ground-truth class. $(1-p_t)^\gamma$ is a modulating factor, with tunable focusing parameter $\gamma\ge0$. And $p_t$ is the model's estimated probability for the ground-truth class. For notational convenience, $p_t$ is defined as follows:
\begin{equation}
    \label{eq5}
    p_t =
    \begin{cases}
        p,&if\ \ \  y = 0\\
        1 - p,&otherwise
    \end{cases}
\end{equation}
in the above $y\in\left\{0,1\right\}$ specifies the ground-truth class. And p is the model's estimated probability for genuine class. During training, when the $p_t$ is small and an example is misclassified, the modulating factor is near 1 and the loss is nearly unaffected. But As $p_t\xrightarrow{}1$, the factor goes to 0 and the loss for well-classified examples is down-weighted. As $\gamma$ increases, the effect of the modulating factor is likewise increased.
\section{Experiment setup}
In this section, we first present the dataset and evaluation metrics that we used in the experiment. Then, we introduce the details of system implementation.
\subsection{Dataset and evaluation metrics}
We use the logical access corpus of ASVspoof2019 Challenge \cite{wang2020asvspoof} to evaluate the performance of our proposed system. As illustrated in table \ref{table:summary_asvspoof2019}, the LA dataset is partitioned into three parts for training, development and evaluation, and each part includes genuine speech and spoofing attacks that generated from different kinds of text to speech (TTS) and voice conversion (VC) algorithms. There exist the same 6 attacks (A01-A06) in the training and development set and 13 attacks (A07-A19) in the evaluation set. It is noted that the evaluation set includes only two known attacks (A16, A19) and other spoofing attacks are unknown (A07-A15, A17, A18).   
We use the equal error rate (EER) and the tandem detection cost function (t-DCF) as the primary metrics to evaluate the proposed anti-spoofing system. The t-DCF takes both the ASV system and spoofing countermeasure errors into consideration. For the ASV system, the lower the t-DCF, the better the system performance. All results reported in this paper are the best of three runs with different random seeds.
\begin{table}
\caption{Summary of the ASVspoof 2019 LA dataset}
\label{table:summary_asvspoof2019}
    \centering
    \resizebox{\linewidth}{!}{
        \begin{tabular}{c|c|c|c}
        \hline\hline
        \multirow{2}{*}{~} & {bonafide} & \multicolumn{2}{c}{Spoof} \\
        \cline{2-4}
        ~ & \#utterance & \#utterance & \#attacks\\
        \hline
        Training & 2580 & 22800 & A01-A06\\
        Development & 2548 & 22296 & A01-A06\\
        Evaluation & 7355 & 63882 & A07-A19\\
        \hline\hline
        \end{tabular}
    }
\end{table}
\subsection{Details of system implementation}
In this study, We do not extract the hand-crafted features from the raw waveform. We directly use the raw waveform as the input of the proposed network to train the model end-to-end. Since the speech data of ASVspoof 2019 is recorded with varying durations, we have to align the speech data, all examples should be truncated or repeated until the duration is the same. And in the experiments, we keep all the speech data of LA dataset with 6 seconds.

We train each model with 50 epochs for LA dataset. The batch size is set to 32. The model with the lowest EER on the development set is chosen to be evaluated. We select AdamW \cite{loshchilov2018decoupled} as the optimizer with an initial learning rate of 0.001, $\beta_1$ and $\beta_2$ are set to 0.9 and 0.999 respectively. Exponential learning rate decay with a multiplicative factor of 0.97 is adopted. We used focal loss as the model's loss function, and the loss weight $\alpha_t$ is decided by the ratio of the spoof and genuine speech in the training dataset. We set $\gamma$ to 2 in the focal loss. Finally, the anti-spoofing system's output is directly adopted as the countermeasure (CM) score.

\begin{table*}
\caption{Breakdown EER (\%) performance of all 13 attacks that exists in the ASVSpoof 2019 LA evaluation set, pooled min t-DCF(P1), and pooled EER (\%, P2). Two state-of-the-art methods and the proposed model are reported. All reported performances are the best of three runs with different random seeds. The best performance for each column is marked in boldface.}
\centering
\resizebox{\linewidth}{!}{
    \begin{tabular}{c|c|c|c|c|c|c|c|c|c|c|c|c|c|cc}
    \hline\hline
    {System} & {A07} & {A08} & {A09} & {A10} & {A11} & {A12} & {A13} & A{14} & {A15} & {A16} & {A17} & {A18} & {A19} & {P1} & {P2 (\%)}\\
    \hline
    Wang et al.\cite{hansen22_interspeech} & 0.91 & 0.29 & 0.02 & 1.39 & 0.31 & 1.30 & 0.08 & 0.18 & 0.61 & \bf{0.24} & 2.11 & 2.25 & 0.97 & 0.0289 & 0.99\\
    Jung et al.\cite{jung2022aasist} & 0.80 & 0.44 & \bf{0.00} & 1.06 & 0.31 & 0.91 & 0.10 & \bf{0.14} & 0.65 & 0.72 & \bf{1.52} & 3.40 & 0.62 & 0.0275 & 0.83\\
    \bf{Ours} & \bf{0.15} & \bf{0.15} & 0.02 & \bf{0.41} & \bf{0.10} & \bf{0.06} & \bf{0.02} & 0.35 & \bf{0.41} & 0.30 & 2.19 & \bf{0.27} & \bf{0.42} & \bf{0.0187} & \bf{0.64}\\
    \hline\hline
    \end{tabular}
}
\end{table*}

\section{Results and analysis}
In this section, we first present the ablation study for the proposed method and then a comparison of our result to the state-of-the-art systems is presented.
\subsection{Ablation study}
To further demonstrate the merit of the proposed channel attention module (MECA), we perform ablation experiments for a such module. As shown in Table 3, without the channel attention module, the proposed model could achieve an equal error rate (EER) of 2.39\% and min-tDCF of 0.0615 in the ASVSpoof 2019 LA evaluation dataset. Integrated with the channel attention module, the equal error rate and the min-tDCF could reduce to 0.64\% and 0.0187 respectively, which demonstrate the effectiveness of the proposed channel attention module (MECA). And all these results are much better than the ASVSpoof 2019 LA baselines.

\begin{table}
\caption{Results for the ablation study of the channel attention module (MECA) in the proposed model and the comparison with ASVSpoof 2019 baseline on the ASVspoof 2019 dataset in terms of min t-DCF and pooled EER.}
\label{table:ablation_meca}
    \centering
    \resizebox{\linewidth}{!}{
        \begin{tabular}{c|c|c|c|c}
        \hline\hline
        \multirow{2}{*}{System} & \multicolumn{2}{c}{Development Set} & \multicolumn{2}{c}{Evaluation Set} \\
        \cline{2-5}
        ~ & min-tDCF & EER(\%) & min-tDCF & EER(\%)\\
        \hline
        CQCC-GMM \cite{todisco2019asvspoof} & 0.0123 & 0.43 & 0.2366 & 9.57\\
        LFCC-GMM \cite{todisco2019asvspoof} & 0.0663 & 2.71 & 0.2116 & 8.09\\
        \hline
        CNBNN & 0.0115 & 0.42 & 0.0615 & 2.39\\
        CNBNN (MECA) & \bf{0.0037} & \bf{0.16} & \bf{0.0187} & \bf{0.64}\\
        \hline\hline
        \end{tabular}
    }
\end{table}

\begin{figure}[ht]
\centerline{\includegraphics[scale=0.26]{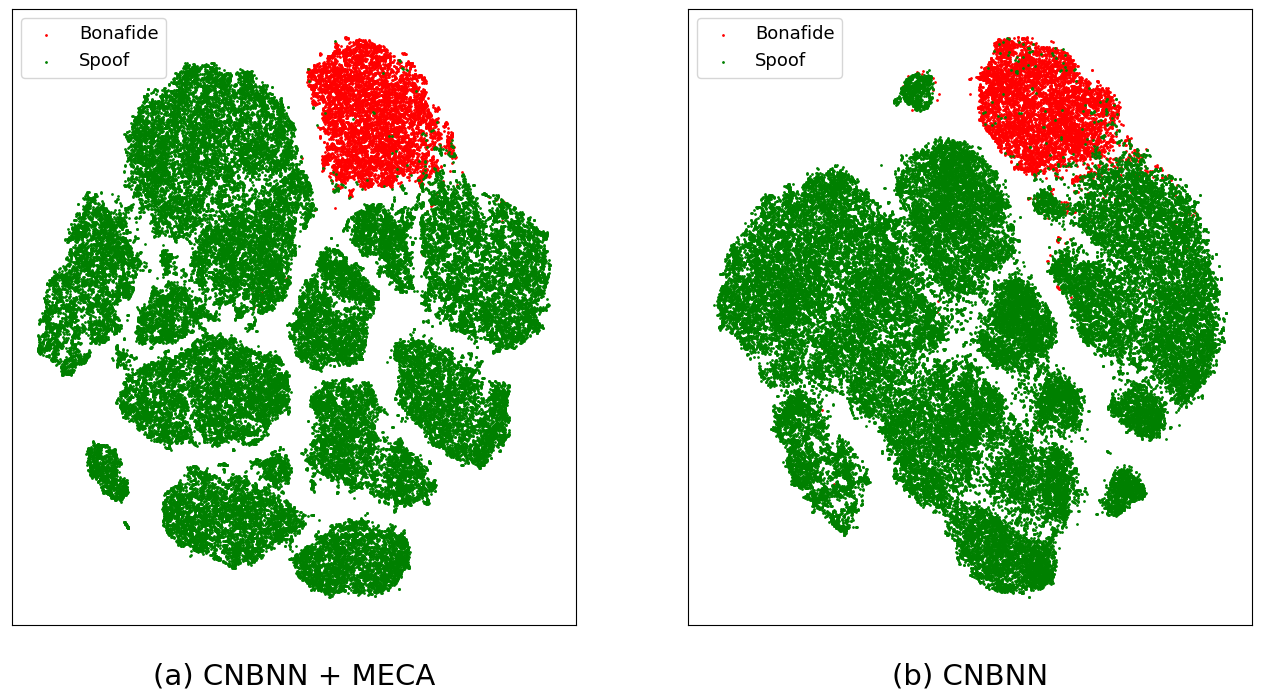}}
\caption{Visualization of t-SNE embeddings from proposed systems in ablation study evaluated on the ASVspoof 2019 LA evaluation dataset. Each color represents whether the speech is genuine (bonafide) or spoofed.}
\label{fig:t-sne}
\end{figure}

As illustrated in Figure \ref{fig:t-sne}, we visualize embeddings from the proposed system evaluated on ASVspoof 2019 LA evaluation dataset by applying t-SNE \cite{van2008visualizing}. We can see that embeddings of the proposed model that use channel attention module are more discriminative and form a well-constructed manifold in embedding space than the system that without the channel attention module. This result shows that the proposed system that using the channel attention module have a better capacity for feature extraction so as to detect the spoof attacks more effectively.

\subsection{Comparison with state-of-the-art systems}
As illustrated in Table 4, we compare our proposed best single system with competing single state-of-the-art systems. These results show that our proposed model could outperform many systems by a large margin. And compare with other single systems, the model that we used for anti-spoofing is still light-weight, it has only 339K parameters in the neural network.

\begin{table}
\caption{Performance on the ASVspoof 2019 evaluation partition in terms of min t-DCF and pooled EER for state-of-the-art systems and our proposed best system.}
\centering
\resizebox{\linewidth}{!}{
    \begin{tabular}{c|c|c|c|c}
        \hline\hline
        {System} & {\#Param} & {Front-end} & {min t-DCF} & {EER(\%)}\\
        \hline
        \bf{Ours} & \bf{339K} & \bf{Raw waveform} & \bf{0.0187} & \bf{0.64}\\
        AASIST \cite{jung2022aasist} & 297K & Raw waveform & 0.0275 & 0.83\\
        RawNet2 + SimAM \cite{hansen22_interspeech} & - & Raw waveform & 0.0289 & 0.99\\
        RawGAT-ST \cite{tak21_asvspoof} & 437K & Raw waveform & 0.0335 & 1.06\\
        SENet \cite{zhang21da_interspeech} & 1,100K & FFT & 0.0368 & 1.14\\
        Non-OFD \cite{choi2022overlapped} & 106K & CQT & - & 1.35\\
        ECAPA-TDNN \cite{9746722} & - & FastAudio-Tri & 0.0451 & 1.54\\
        Res-TSSDNet \cite{hua2021towards} & 350K & Raw waveform & 0.0481 & 1.64\\
        GMM-ResNet \cite{9746163} & - & LFCC & 0.0498 & 1.80\\
        Raw PCDARTS \cite{ge21_asvspoof} & 24,480K & Raw waveform & 0.0517 & 1.77\\
        MCG-Res2Net50 \cite{li21o_interspeech} & 960K & CQT & 0.0520 & 1.78\\
        LCNN-LSTM-sum \cite{wang21fa_interspeech} & 276K & LFCC & 0.0524 & 1.92\\
        Resnet18-OC-softmax \cite{zhang2021one} & - & LFCC & 0.0590 & 2.19\\
        SE-Res2Net50 \cite{li2021replay} & 920K & CQT & 0.0743 & 2.50\\
        PC-DARTS \cite{ge21c_interspeech} & 7510K & LFCC & 0.0914 & 4.96\\
        \hline\hline
    \end{tabular}
}
\end{table}

We also perform the comparison with other state-of-the-art systems for the 13 attacks in the evaluation dataset. As illustrated in Table 2, our proposed best single system could outperform the state-of-the-art AASIST system on 9 attacks (A07, A08, A10-A13, A15, A18 and A19), especially, the absolute EER reducion on A18 attack is 3.13\%, which is reduce by a large margin.
\section{Conclusions}
This work proposes a new end-to-end anti-spoofing system. We revise the ConvNeXt network architecture and then integrate a channel attention module that is modified from the original ECA layer with it to further enhance the model's anti-spoofing performance. When using the focal loss, the proposed best single system could achieve state-of-the-art performance on LA scenarios. In the future, we will apply the proposed system to other speech tasks.

\footnotesize
\bibliographystyle{IEEEbib}
\bibliography{icme2022template}

\end{document}